\documentclass[aps,prb,twocolumn,showpacs]{revtex4}
\usepackage{graphicx}

\begin{document}

\title{Synthesis, structural and transport properties of the hole-doped Superconductor Pr$_{1-x}$Sr$_x$FeAsO }

\author{Gang Mu, Bin Zeng, Xiyu Zhu, Fei Han, Peng Cheng, Bing Shen, and Hai-Hu Wen}\email{hhwen@aphy.iphy.ac.cn }

\affiliation{National Laboratory for Superconductivity, Institute of
Physics and Beijing National Laboratory for Condensed Matter
Physics, Chinese Academy of Sciences, P. O. Box 603, Beijing 100190,
People's Republic of China}

\begin{abstract}
Superconductivity was achieved in PrFeAsO by partially substituting
Pr$^{3+}$ with Sr$^{2+}$. The electrical transport properties and
structure of this new superconductor Pr$_{1-x}$Sr$_x$FeAsO at
different doping levels (x = 0.05$\sim$ 0.25) were investigated
systematically. It was found that the lattice constants ($a$-axis
and $c$-axis) increase monotonously with Sr or hole concentration.
The superconducting transition temperature at about 16.3 K (95\%
$\rho_n$) was observed around the doping level of 0.20$\sim$ 0.25. A
detailed investigation was carried out in the sample with doping
level of x = 0.25. The domination of hole-like charge carriers in
this material was confirmed by Hall effect measurements. The
magnetoresistance (MR) behavior can be well described by a simple
two-band model. The upper critical field of the sample with $T_c$ =
16.3 K (x = 0.25) was estimated to be beyond 45 Tesla. Our results
suggest that the hole-doped samples may have higher upper critical
fields comparing to the electron-doped ones, due to the higher
quasi-particle density of states at the Fermi level.
\end{abstract}

 \maketitle

\section{Introduction}

The discovery of superconductivity at 26$\;$K in FeAs-based layered
quaternary compound LaFeAsO$_{1-x}$F$_x$ has attracted great
interests in the fields of condensed matter physics and material
sciences.\cite{Kamihara2008,WenAdvMat2008} The family of the
FeAs-based superconductors has been extended rapidly and it can be
divided into three categories. The first category has the general
formula of REFeAsO where RE stands for the rare earth elements and
is abbreviated as the FeAs-1111 phase.\cite{Rainer} The second class
is formulated as (Ba, Sr)$_{1-x}$K$_x$Fe$_2$As$_2$ which is denoted
as FeAs-122 for simplicity.\cite{Rotter,CWCh} The third type
Li$_x$FeAs has an infinite layered structure (denoted as FeAs-
111).\cite{LiFeAs,LiFeAsChu,LiFeAsUK} As for the FeAs-1111 phase,
most of the discovered superconductors are characterized as electron
doped ones and the superconducting transition temperature has been
quickly raised to $T_c$ = 55$\sim$ 56 K via replacing lanthanum with
other rare earth elements.\cite{XHCh,NLW,Pr52K,RenZA55K,CP,WangC}
About the hole-doped side, however, since the first hole-doped
superconductor La$_{1-x}$Sr$_{x}$FeAsO with $T_c \approx$ 25 K was
discovered,\cite{WenEPL,LaSr2} only the Nd-based system
Nd$_{1-x}$Sr$_x$FeAsO with $T_c$ = 13.5 K was reported.\cite{NdSr}
Obviously, there is an extensive space to explore more
superconductors in the hole-doped side based on the FeAs-1111 phase
and to further extend the family of the FeAs-based superconductors.
And it is also significant to investigate the basic physical
properties of the hole-doped system based on the FeAs-1111 phase.

In this paper we report a new route to easily synthesize the
hole-doped superconductors based on the FeAs-1111 phase, Sr doped
Pr$_{1-x}$Sr$_x$FeAsO, with the maximum superconducting transition
temperature of 16.3 K. We carried out a systematic study on the
evolution of the superconductivity and the lattice constants with
the content of Sr or hole concentration in the system of
Pr$_{1-x}$Sr$_{x}$FeAsO. We found that the $a$-axis and $c$-axis
lattice constants increase monotonously with doped concentration of
Sr or hole numbers. The physical properties of a selected sample
with x = 0.25 were investigated in depth. The conducting charge
carriers in this sample were characterized to be hole type by the
Hall effect measurements. And it is found that the MR data show a
good two-band behavior. We also estimated the upper critical field
of the same sample  based on the Ginzburg-Landau theory as well as
the Werthamer-Helfand-Hohenberg (WHH) formula.\cite{WHH} It is
suggested that the upper critical fields in the hole-doped samples
may be higher than that in the electron-doped ones.

\section{Experimental Details}

The Pr$_{1-x}$Sr$_x$FeAsO samples were prepared using a two-step
solid state reaction method. In the first step, PrAs and SrAs were
prepared by reacting Pr flakes (purity 99.99\%), Sr flakes (purity
99.9\%) and As grains (purity 99.99\%) at 500 $^o$C for 8 hours and
then 700 $^o$C for 16 hours. They were sealed in an evacuated quartz
tube when reacting. Then the resultant precursors were thoroughly
grounded together with Fe powder (purity 99.95\%) and Fe$_2$O$_3$
powder (purity 99.5\%) in stoichiometry as given by the formula
Pr$_{1-x}$Sr$_x$FeAsO. All the weighing and mixing procedures were
performed in a glove box with a protective argon atmosphere. Then
the mixtures were pressed into pellets and sealed in a quartz tube
with an atmosphere of 20\% Ar. The materials were heated up to 1150
$^o$C with a rate of 120 $^o$C/hr and maintained for 60 hours. Then
a cooling procedure was followed. It is important to note that the
use of SrAs as the starting material, instead of using SrCO$_3$ or
SrO, is very essential to synthesize the high quality samples, and
to much suppress the secondary impurity phases. This new route makes
our hole-doped samples easily reproduced.

The X-ray diffraction (XRD) measurements of our samples were carried
out by a $Mac$-$Science$ MXP18A-HF equipment with $\theta$-$
2\theta$ scan. The dc magnetization measurements were done with a
superconducting quantum interference device (Quantum Design, SQUID,
MPMS7) and the ac susceptibility of the samples were measured on the
Maglab-12T (Oxford) with an ac field of 0.1 Oe and a frequency of
333 Hz. The resistance and Hall effect measurements were done using
a six-probe technique on the Quantum Design instrument physical
property measurement system (PPMS) with magnetic fields up to 9 T.
The temperature stabilization was better than 0.1\% and the
resolution of the voltmeter was better than 10 nV.

\begin{figure}
\includegraphics[width=8cm]{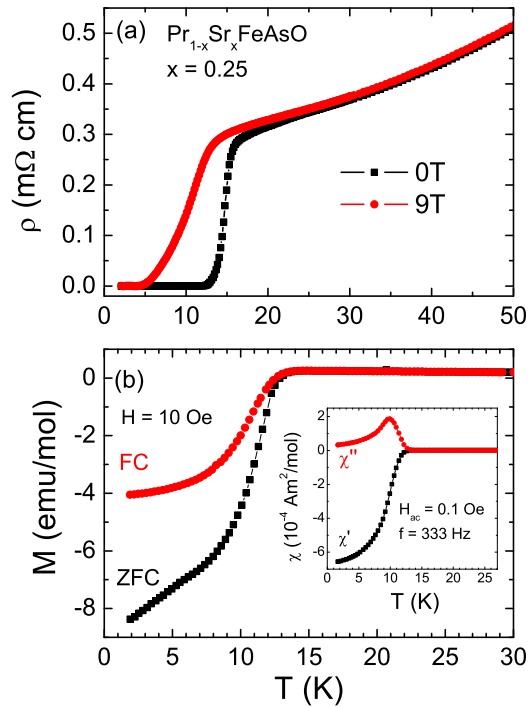}
\caption{(Color online) (a) Temperature dependence of resistivity
for the Pr$_{0.75}$Sr$_{0.25}$FeAsO sample under two different
magnetic fields 0 T and 9 T near the superconducting transition. A
rather sharp transition can be seen at zero field. (b) Temperature
dependence of dc magnetization for the zero field cooling (ZFC) and
field cooling (FC) process at H = 10 Oe. The inset shows the ac
susceptibility of the same sample measured with H$_{ac}$ = 0.1 Oe, f
= 333 Hz. } \label{fig1}
\end{figure}

\begin{figure}
\includegraphics[width=9cm]{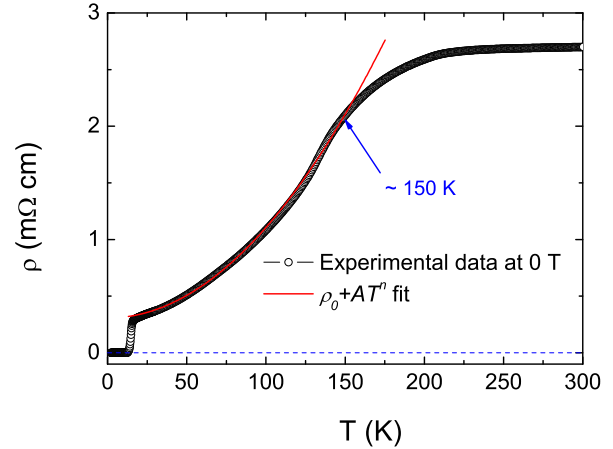}
\caption{(Color online) The resistivity curve at 0 T in the
temperature region up to 300 K for the sample with x = 0.25. A
flattening of resistivity in high temperature region is obvious,
which seems to be a common feature of the hole-doped FeAs-based
superconductors\cite{WenEPL,LaSr2,Rotter}. The red solid line shows
the fit below about 150 K using the formula $\rho=\rho_0+AT^n$. }
\label{fig2}
\end{figure}

\section{Experimental data and discussion}

\subsection{Resistive and diamagnetic transition}

In Fig.1 (a) we present a typical set of resistive data for the
sample Pr$_{1-x}$Sr$_x$FeAsO with x = 0.25 under 0 T and 9 T near
the superconducting transition. One can see that the resistivity
transition at zero field is rather sharp indicating the quite high
quality of our sample, and the onset transition temperature is about
16.3 K taking a criterion of 95\% $\rho_n$. A magnetic field of 9 T
only depresses the onset transition temperature about 2.5 K but
makes the superconducting transition broader. The former behavior
may indicate a rather high critical field in our sample, while the
latter reflected the weak link between the grains.\cite{XYZ} Fig.1
(b) shows the zero field cooled and also the field cooled dc
magnetization of the same sample at 10 Oe. And the diamagnetic
transition measured with ac susceptibility technique is shown in the
inset of Fig.1 (b). A rough estimate from the diamagnetic signal
shows that the superconducting volume fraction of the present sample
is beyond 50\%, confirming the bulk superconductivity in our
samples. The onset critical temperature by magnetic measurements is
roughly corresponding to the zero resistivity temperature.

Shown in Fig. 2 is the temperature dependence of resistivity under
zero field up to 300 K for the same sample as shown in Fig. 1. The
resistivity data in the normal state were fitted using the formula
\begin{equation}
\rho=\rho_0+AT^n,\label{eq:1}
\end{equation}
as we had done in the F-doped LaFeAsO system.\cite{XYZ} As
represented by the red solid line in Fig. 2, the data below about
150 K can be roughly fitted with the fitting parameters $\rho_0 =
0.306$ m$\Omega$ cm and n = 2.000. The fine quadratic dependent
behavior of the resistivity, which is consistent with the prediction
of the Fermi-liquid theory, may suggest a rather strong scattering
between electrons in the present system in the low temperature
region. In high temperature region above about 150 K, however, a
flattening of resistivity was observed clearly. The similar behavior
has been observed in other hole-doped FeAs-1111 systems
La$_{1-x}$Sr$_x$FeAsO and
Nd$_{1-x}$Sr$_x$FeAsO,\cite{WenEPL,LaSr2,NdSr} and also in the
FeAs-122 system (Ba, Sr)$_{1-x}$K$_x$Fe$_2$As$_2$.\cite{Rotter} We
have pointed out that this behavior may be a common feature of the
hole-doped FeAs-based superconductors\cite{LaSr2}.
\begin{figure}
\includegraphics[width=9cm]{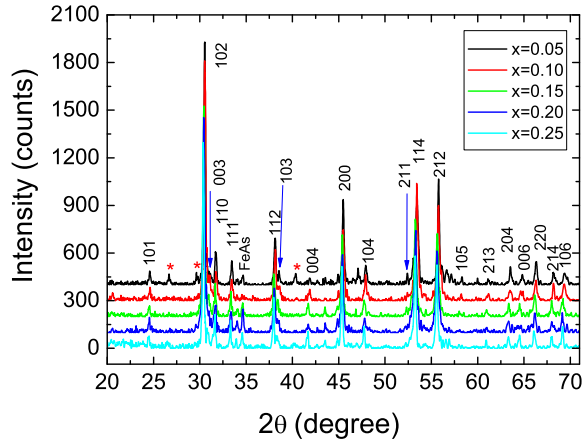}
\caption{(Color online) X-ray diffraction patterns for
Pr$_{1-x}$Sr$_x$FeAsO samples with different doping levels: x =
0.05, 0.10, 0.15, 0.20 and 0.25. One can see that all the main peaks
can be indexed to the tetragonal ZrCuSiAs-type structure. Small
amount of impurity phases were denoted by red asterisks. }
\label{fig3}
\end{figure}

\subsection{Doping dependence of lattice constants and superconducting properties}

The XRD patterns for the samples with the nominal doping levels of
0.05$\sim$ 0.25 are shown in Fig. 3. Some small peaks from small
amount of FeAs impurity phase which were denoted by red asterisk can
still be seen, and the sample with x = 0.05 has a bit more impurity
phase FeAs than other samples. However, it is clear that all the
main peaks can be indexed to the FeAs-1111 phase with the tetragonal
ZrCuSiAs-type structure. By having a closer scrutiny one can find
that the diffraction peaks shift slightly to the low-angle side when
more strontium are doped into the samples, suggesting an expanding
effect of the lattice constants. The similar behavior has been
observed in the Sr-doped LaFeAsO system previously.\cite{LaSr2} By
using the software of Powder-X\cite{DongC}, we took a general fit to
the XRD data of each sample and determined the lattice constants.
The doping dependence of the lattice constants calculated from the
XRD data, along with the data of the undoped and 11\% F-doped
PrFeAsO samples\cite{Pr52K}, are presented in Fig. 4. It is clear
that both the $a$-axis and $c$-axis lattice constants expand
monotonously from 11\% F-doped PrFeAsO to the Sr-doped samples. This
indicates that the strontium atoms go into the crystal lattice of
the PrFeAsO system because the radii of Sr$^{2+}$ (1.12 $\AA$) is
larger than that of Pr$^{3+}$ (1.01 $\AA$). It is worth noting that
the extent of the lattice expanding is appreciably smaller than that
in the Sr-doped LaFeAsO system, where the maximum onset transition
temperature can be as high as 26 K.\cite{WenEPL,LaSr2} In some
cases, we see a slight drop of resistivity at temperatures as high
as 28 K. So we believe that a further increase in $T_c$ is possible
if more strontium can be chemically doped into this system, like in
the case of Nd$_{1-x}$Sr$_x$FeAsO.\cite{NdSr}
\begin{figure}
\includegraphics[width=8cm]{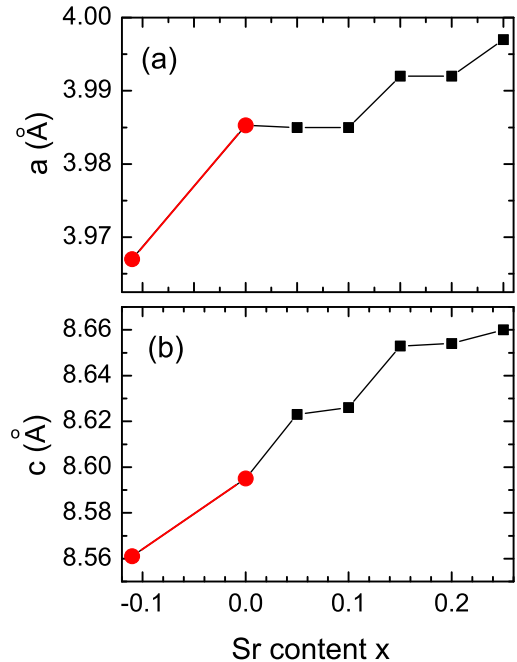}
\caption{(Color online)  Doping dependence of (a) a-axis lattice
constant; (b) c-axis lattice constant. The black filled squares
represent data from our measurements. For electron-doped side, the
variable "x" represents the fluorine concentration. The data of the
undoped and F-doped cases (red filled circles) are taken from the
work of Ren et al.\cite{Pr52K} One can see that the lattice constant
roughly show a monotonous variation versus doping from electron
doped region to hole-doped one. } \label{fig4}
\end{figure}

In Fig. 5 we show the resistivity data of our samples made at
various nominal doping levels of Sr ranging from x = 0.05 to 0.25. A
clear but rounded resistivity anomaly can be seen around
155$\sim$175 K when the doping level is 5\%. And a tiny resistivity
drop at about 6 K which may be induced by the antiferromagnetic
ordering of Pr$^{3+}$ ions or superconductivity can be observed. At
this time it is difficult to discriminate between the two scenarios
since the magnitude of the resistivity drop is quite small. At the
doping levels of 0.10$\sim$ 0.25, resistivity anomaly in high
temperature regime is suppressed gradually and it eventually evolves
into a flattening behavior at high doping levels. Also the magnitude
of resistivity reduces obviously compared with the sample with 5\%
doping, suggesting that more and more conducting charge carriers
were introduced into the samples. At the same time, the
superconductivity emerges with doping and becomes optimal with an
onset transition temperature of 16.3 K when the doping level is x =
0.20$\sim$ 0.25. It is worth noting that the resistivity at high
temperatures at a high doping level of holes behaves in a different
way compared with that in the electron-doped
samples\cite{Pr52K,SDW,YJia} where the resistivity anomaly is
suppressed completely and the resistivity always presents a metallic
behavior in that regime. While in the hole-doped samples, the
resistivity anomaly is smeared much slower.\cite{WenEPL,LaSr2,NdSr}
In the Nd$_{1-x}$Sr$_x$FeAsO system,\cite{NdSr} for example, it is
found that the structural transition from tetragonal to orthorhombic
occurs even in the sample with superconductivity. Doping more holes
may lead to stronger suppression to the structural transition as
well as the SDW order, this may leave a potential space to increase
the superconducting transition temperature in the hole-doped
FeAs-1111 phase.

\begin{figure}
\includegraphics[width=8.5cm]{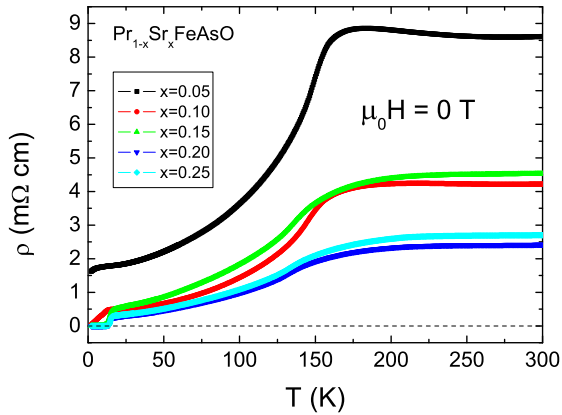}
\caption{(Color online) Temperature dependence of resistivity of
samples Pr$_{1-x}$Sr$_{x}$FeAsO with x = 0.05, 0.10, 0.15, 0.20,
0.25. One can see that the resistivity anomaly is suppressed
gradually by doping more Sr into the system. The maximum onset
transition temperature appears around the nominal doping level of
0.20$\sim$ 0.25. } \label{fig5}
\end{figure}

\begin{figure}
\includegraphics[width=8cm]{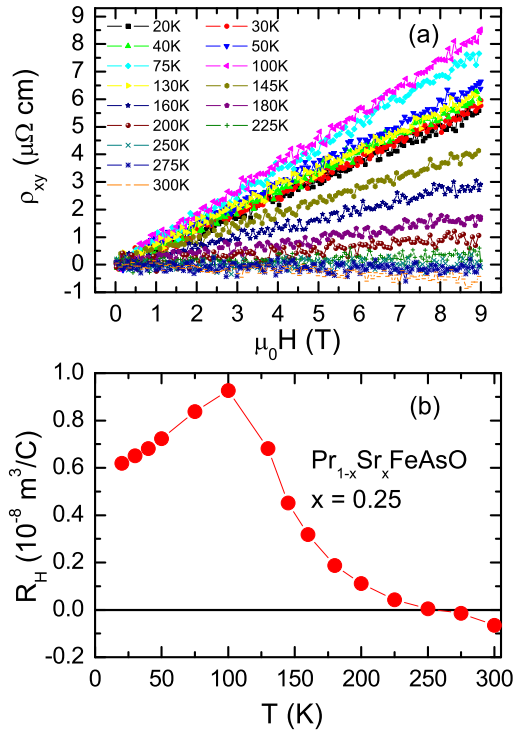}
\caption{(Color online) Hall effect measurements for one sample
Pr$_{0.75}$Sr$_{0.25}$FeAsO. (a) Hall resistivity $\rho_{xy}$ versus
the magnetic field $\mu_0$H at different temperatures. (b)
Temperature dependence of the Hall coefficient $R_\mathrm{H}$. A
huge hump appears in the intermediate temperature region, which
seems to be a common feature for the hole-doped FeAs-based
superconductors. } \label{fig6}
\end{figure}

\subsection{Hall effect}

It is known that for a conventional metal with Fermi liquid feature,
the Hall coefficient is almost independent of temperature. However,
this situation is changed for a multiband material\cite{HY} or a
sample with non-Fermi liquid behavior, such as the cuprate
superconductors.\cite{Ong} To get more information about the
conducting carriers, we measured the Hall effect of the sample with
x = 0.25 (the same one as shown in Fig. 1). Fig. 6(a) shows the
magnetic field dependence of Hall resistivity ($\rho_{xy}$) at
different temperatures. In the experiment $\rho_{xy}$ was taken as
$\rho_{xy}$ = [$\rho$(+H) - $\rho$(-H)]/2 at each point to eliminate
the effect of the misaligned Hall electrodes. It is clear that all
curves in Fig. 6(a) have good linearity versus the magnetic field.
Moreover, $\rho_{xy}$ is positive at all temperatures below 250 K
giving a positive Hall coefficient $R_H = \rho_{xy}/H$, which
actually indicates that hole type charge carriers dominate the
conduction in the present sample.

The temperature dependence of $R_H$ is shown in Fig. 6(b). Very
similar to that observed in La$_{1-x}$Sr$_{x}$FeAsO
samples,\cite{WenEPL,LaSr2} the Hall coefficient $R_H$ reveals a
huge hump in the intermediate temperature regime and the value of
$R_H$ decreases down to zero at about 250 K, then it becomes
slightly negative above that temperature. Here we employ a simple
two-band scenario with different types of carriers to interpret this
behavior. We have known that for a two-band system in the low-field
limit, the Hall coefficient $R_H$ can be written as
\begin{equation}
R_H=\frac{\sigma_1^2 R_1+\sigma_2^2
R_2}{(\sigma_1+\sigma_2)^2},\label{eq:2}
\end{equation}
where $\sigma_i$ (i = 1, 2) is the conductance for different types
of charge carriers in different bands, and $R_i = -1/n_ie$
represents the Hall coefficient for each type of carriers separately
with $n_i$ the concentration of the charge carriers for the
different bands. We attribute the strong temperature dependence of
$R_H$ in the present system to the complicated variation of the
conductance $\sigma_i$ with temperature, which reflects mainly the
temperature dependent behavior of the scattering relaxation time.
And the sign-changing effect of $R_H$ may indicate the presence of
two different types of charge carriers (electron and hole type) in
the present system. The conductance $\sigma_i$ of electron-like and
hole-like carriers may vary differently with temperature, and the
electron-like carriers become dominant when the temperature is
higher than about 250 K.

This simple two-band model is consistent with the MR data (will be
addressed in the next section). However, if there are two types of
carriers in two bands, we may expect $\rho_{xy}$ to be quadratic in
magnetic field. The linear behavior in the $\rho_{xy} \sim H$ curve
shown in Fig. 6(a) seems to be not agree with this scenario. So
further measurements with higher magnetic fields are required.

\begin{figure}
\includegraphics[width=8cm]{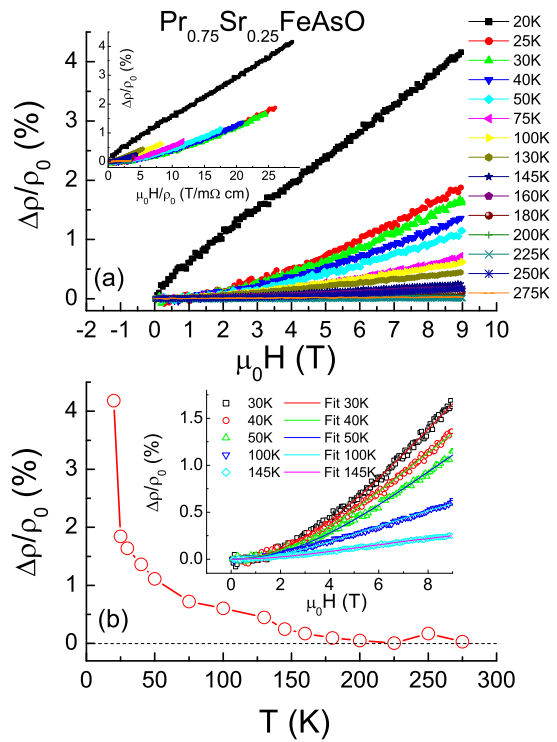}
\caption{(Color online) (a) Field dependence of MR
$\Delta\rho/\rho_0$ at different temperatures for the sample with x
= 0.25. The inset shows Kohler plot of the same sample. It is clear
that Kohler¡¯s rule is not obeyed. (b) Temperature dependence of MR
for the same sample determined at 9 T. Shown in the inset is the
theoretical fit to the field dependent MR data using a two-band
model (see text). } \label{fig7}
\end{figure}

\begin{figure}
\includegraphics[width=8.5cm]{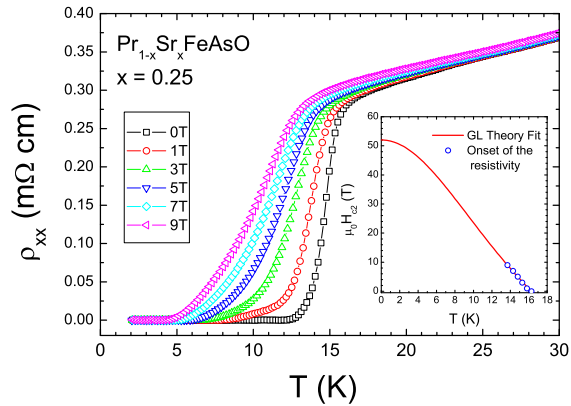}
\caption{(Color online) Temperature dependence of resistivity for
the Pr$_{0.75}$Sr$_{0.25}$FeAsO sample under different magnetic
fields. The onset transition temperature defined by 95\%$\rho_n$
shifts with the magnetic field slowly. The inset shows the phase
diagram derived from the resistive transition curves. The onset
transition point is presented by the blue open circles. The red
solid line shows the theoretical curve based on the GL theory (Eq.
5).} \label{fig8}
\end{figure}

\subsection{Magnetoresistance}

Magnetoresistance is a very powerful tool to investigate the
electronic scattering process and the information about the Fermi
surface.\cite{LiQ,HY} Field dependence of MR, for the sample with x
= 0.25 at different temperatures is shown in the main frame of Fig.
7(a). Here MR was expressed as $\Delta\rho/\rho_0 =
[\rho(H)-\rho_0]/\rho_0$, where $\rho(H)$ and $\rho_0$ represent the
longitudinal resistivity at a magnetic field H and that at zero
field, respectively. One can see that the curve obtained at 20 K
reveals a rather different feature compared with the data at
temperatures above 25 K. Also the magnitude of MR (obtained at 9 T)
at 20 K reached 2 times of that at 25 K giving a sharp drop in the
$\Delta\rho/\rho_0 \sim T$ curve when increasing the temperature as
revealed in the main frame of Fig. 7. We attribute this behavior to
the presence of fluctuant superconductivity in the low temperature
region ($\sim$16.3 K-20 K). At temperatures higher than 150 K, the
value of MR vanishes to zero gradually.

As for the case in the moderate temperature region, the
$\Delta\rho/\rho_0 \sim H$ curve reveals a clear nonlinear behavior.
In the inset of Fig. 7(b) we present the data at five typical
temperatures in this region. These data were then fitted based on a
simple two-band model which gave the following formula
\begin{equation}
\frac{\Delta\rho}{\rho_0}=\frac{(\mu_0 H)^2}{\alpha+\beta\times
(\mu_0 H)^2},\label{eq:3}
\end{equation}
with $\alpha$ and $\beta$ the fitting parameters which were related
to the conductances and mobilities for the charge carriers in two
bands. The fitting results were shown by the solid lines in the
inset of Fig. 7(b). It is clear that Eq. 3 can describe our data
quite well. This argument can be further confirmed by seeing about
the situation of the so-called Kohler plot. The semiclassical
transport theory has predicted that the Kohler rule, which can be
written as
\begin{equation}
\frac{\Delta\rho}{\rho_0}= F(\frac{\mu_0 H}{\rho_0}),\label{eq:4}
\end{equation}
will be held if only one isotropic relaxation time is present in a
single-band solid-state system.\cite{KL} Eq. (4) means that the
$\Delta\rho/\rho_0$ vs $\mu_0 H/\rho_0$ curves for different
temperatures should be scaled to a universal curve if the Kohler
rule is obeyed. The scaling based on the Kohler plot for the present
sample is revealed in the inset of Fig. 7(a). An obvious violation
of the Kohler rule can be seen on this plot. We attribute this
behavior to the presence of a multi-band effect in the present
system.

Actually, theoretical researches and angle-resolved photoemission
spectroscopic (ARPES) studies have shown a rather complicated Fermi
surface and energy-band structure in the FeAs-based superconductors.
Our data from measuring the Hall effect and MR are consistent with
these conclusions.

\subsection{Upper critical field}

Finally, we attempted to estimate the upper critical field of the
sample with x = 0.25 from the resistivity data. Temperature
dependence of resistivity under different magnetic fields is shown
in the main frame of Fig. 8. Similar to that found in the F-doped
LaFeAsO polycrystalline samples,\cite{XYZ,WangNL2} the onset
transition point, which reflects mainly the upper critical field in
the configuration of H$\|$ab-plane, shifts more slowly than the zero
resistivity point to low temperatures under fields. We take a
criterion of 95\%$\rho_n$ to determine the onset transition points
under different fields, which are presented by the blue open circles
in the inset of Fig. 6. From these data we can roughly estimate the
upper critical field of this sample based on the Ginzburg-Landau
(GL) theory. The following equation have been extract from the GL
theory and used successfully on other samples:\cite{XYZ,WangNL2,NTS}

\begin{equation}
H_{c2}(T)=H_{c2}(0)\frac{1-t^2}{1+t^2},\label{eq:5}
\end{equation}
where $t=T/T_c$ is the reduced temperature and $H_{c2}(0)$ is the
upper critical field at zero temperature. Taking $T_c$= 16.3 K and
$H_{c2}(0)$ as the adjustable parameter, the measured data in the
inset of Fig. 8 were then fitted using Eq. 5. One can see a quite
good fit in the inset of Fig. 8 as revealed by the red solid line.
The zero temperature upper critical field was determined to be
$H_{c2}(0) \approx$ 52 T from the fitting process. Actually one can
also determine the slope of $H_{c2}(T)$ near $T_c$, which is found
to be about -4.0 T/K in the present sample. By using the WHH
formula\cite{WHH} the value of zero temperature upper critical field
$H_{c2}(0)$ can be estimated through:
\begin{equation}
H_{c2}(0)=-0.693T_c(\frac{dH_{c2}}{dT})_{T=T_c}. \label{eq:6}
\end{equation}
Taking $T_c$= 16.3 K, we get $H_{c2}(0) \approx 45.1$ T. Regarding
the relatively low value of $T_c$=16.3 K in the present sample, this
value of upper critical field $H_{c2}(0)$ is actually quite high.
The slope of $dH_{c2}(T)/dT|_{T_c}$ in the hole-doped sample is
clearly larger than that in the electron-doped samples. In the
hole-doped La$_{1-x}$Sr$_x$FeAsO superconducting samples, we also
found much larger $dH_{c2}(T)/dT|_{T_c}$ when comparing it with the
F-doped LaFeAsO sample.\cite{XYZ,WangNL2} This may be understood as
due to the higher quasiparticle density of states (DOS) near the
Fermi level in the hole-doped samples. In a dirty type-II
superconductor, it was predicted\cite{JaffePRB1989} that
$-dH_{c2}/dT|_{T_c}\propto \gamma_n$ with $\gamma_n$ the normal
state specific heat coefficient which is proportional to the DOS at
$E_F$. It is thus reasonable to ascribe the higher value of
$dH_{c2}/dT|_{T_c}$ to higher DOS in the hole-doped samples.
Theoretical calculations do show that the DOS in the hole-doped side
is larger than that in the electron-doped
samples.\cite{SinghPRB,LuZY} The measurements on lower critical
fields\cite{RenC} and specific heat\cite{MuG} in
Ba$_{0.6}$K$_{0.4}$Fe$_2$As$_2$ reveal that the superfluid density
and the normal state DOS is about 5 to 10 times larger than that in
the F-doped REFeAsO system. If the normal state DOS is really larger
in the hole-doped systems, higher upper critical fields may be
achieved in the hole-doped FeAs-1111 samples provided that the
superconducting transition temperature can be improved to the same
scale.

\section{Concluding remarks}

In summary, bulk superconductivity was achieved by substituting
Pr$^{3+}$ with Sr$^{2+}$ in PrFeAsO system. A systematic evolution
of superconductivity and the lattice constants with doping in hole
doped Pr$_{1-x}$Sr$_x$FeAsO was discovered. By doping more Sr into
the parent phase PrFeAsO, the anomaly of resistivity at about 165 K
is suppressed gradually and the superconductivity eventually sets
in. The $a$-axis and $c$-axis lattice constants increase
monotonously with Sr concentration. The maximum superconducting
transition temperature $T_c$ = 16.3 K is found to appear around the
nominal doping level x = 0.20$\sim$ 0.25. The positive Hall
coefficient $R_H$ in a wide temperature range suggests that the hole
type charge carriers dominate the conduction in this system. The
strong temperature dependence and sign-changing effect of $R_H$ were
attributed to a multi-band effect and have been interpreted based on
a simple two-band model with different types of charge carriers.
This argument was further confirmed by the nonlinear field
dependence of MR and the violation of the Kohler rule.
Interestingly, the slope of the upper critical magnetic field vs.
temperature near $T_c$ seems to be much higher than that of the
electron-doped samples. This is attributed to the higher DOS in the
hole-doped samples than in the electron-doped ones. This may provide
a new way to enhance the upper critical field in the hole-doped
FeAs-1111 superconductors.

\begin{acknowledgments}
We acknowledge the help of XRD experiments from L. H. Yang and H.
Chen. This work is supported by the Natural Science Foundation of
China, the Ministry of Science and Technology of China (973 project:
2006CB01000, 2006CB921802), the Knowledge Innovation Project of
Chinese Academy of Sciences (ITSNEM).
\end{acknowledgments}

\end{document}